\documentclass[runningheads]{comsis2}

\def\journalissue{Computer Science and Information Systems 00(0):0000--0000}
\def\paperidnum{https://doi.org/10.2298/CSIS123456789X}
\usepackage{picins}
\usepackage{hyperref}
\usepackage{verbatim}
\usepackage{graphicx}
\usepackage{color}
\usepackage{cite}
\usepackage{booktabs}
\usepackage{extarrows}
\usepackage{float}
\usepackage[ruled,vlined]{algorithm2e}
	\usepackage{algpseudocode}
	\usepackage{amsmath}
\title{Fabric-GC: A Blockchain-based Gantt Chart System for Cross-organizational Project Management}

\titlerunning{Fabric-GC}

\author{
	Dun Li \inst{1}\textsuperscript{\dag}                \and 
	Dezhi Han \inst{1}\textsuperscript{\dag}	           \and
	Benhui Xia  \inst{1}\textsuperscript{\dag}           \and
	Tien-Hsiung Weng  \inst{2}\textsuperscript{*}
 \and
	Arcangelo Castiglione  \inst{3}\and
	Kuan-Ching Li \inst{2}\textsuperscript{*}
}

\authorrunning{Dun Li et al.}
\institute{College of Information Engineering,
	Shanghai Maritime University\\
	201306, Shanghai, China\\
	\email{\{lidunshmtu@outlook.com}
	\and
	Dept.\ of Computer Science and Information Engr. (CSIE), Providence University\\
	43301, Taichung, Taiwan\\
	\email{kuancli@pu.edu.tw}
	\and
	Department of Computer Science
	University of Salerno, Fisciano\\
	84084, SA, Italy\\
	\email{arcastiglione@unisa.it}
	\\
	\textit{\textsuperscript{*}corresponding authors}
	\\
	\textit{\textsuperscript{\dag} authors contribute equally to this work}
}

\begin{document}

\maketitle

\begin{abstract}
Large-scale production is always associated with more and more development and interaction among peers, and many fields achieve higher economic benefits through project cooperation. However, project managers in the traditional centralized approach cannot rearrange their activities to cross-organizational project management. Thanks to its characteristics, the Blockchain can represent a valid solution to the problems mentioned above. In this article, we propose Fabric-GC, a Blockchain-based Gantt chart system. Fabric-GC enables to realize secure and effective cross-organizational cooperation for project management, providing access control to multiple parties for project visualization. 
Compared with other solutions, the proposed system is versatile, as it can be applied to project management in different fields and achieve effective and agile scheduling. Experimental results show that Fabric-GC achieves stable performance in large-scale request and processing distributed environments, where the data synchronization speed of the consortium chain reached four times faster than a public chain, achieving faster data consistency.

\vspace{6pt}\textbf{Keywords:} 
Cross-organizational secure cooperation,  Blockchain, Gantt chart, Project management,  Hyperledger fabric, Data sharing
\end{abstract}

\section{Introduction}
\label{intro}
Project Management (PM) is an activity carried out by project managers who plan, organize, direct, coordinate, control, and evaluate projects through scientific and management activities to reach the project objectives. Due to limited available resources (i.e., time, budget, labor), there are many constraints on the activities and development that affect the overall schedule of PM. Due to the reasons mentioned above, project management can coordinate and guide project implementation under constraints and limitations, reducing complexity and operational costs and improving the efficiency of project implementation. 
Traditional project management systems typically rely on the \emph{Storage-Business-Interface} triad. \emph{Storage} provides permanent storage and ready access to data, \emph{Business} includes all functional modules of the management system, and \emph{Interface} converts data into meaningful management models through visualization \cite{7053283}. To date, the \emph{Gantt} chart is one of the most commonly used project management tools, as it divides the entire project into smaller portions of tasks sequenced under specific rules (e.g., time). In this way, project managers can track the execution status of each task under a given schedule and monitor the completion level of the entire project. This tracking enables the evaluation of the entire project resource budget, optimizing the completion time schedule to make adjustments to the execution plan, and most importantly, making the right decisions.

With the further advancement of science and technology and the development of productivity, multi-party project cooperation is a standard practice, widely used in scientific R \& D, industrial production, software development, supply chain \cite{9047881} among several other fields. Indeed, the collaboration between organizations and individuals with different technologies enlarges the rate for the success of complex projects \cite{7851142}. Nevertheless, cross-organizational projects pose difficulties for project managers in managing task scheduling and progress feedback that relies on timely information sharing \cite{YANG2020103276}. The independence and heterogeneity among participating organizations may turn data sharing difficult. Besides, traditional data sharing relies on third-party organizations (e.g., cloud, specialized service provider, transcription services, call center services, consulting), and therefore, the privacy and security of data cannot be guaranteed \cite{9142202,p5,p4, li2021blockchain}. In fact, although in general, the sharing of information is of great benefit and provides several advantages for all the entities involved, however, these entities may not trust each other, or even worse, they may compete with each other.
Consequently, in the context of cross-organizational project management, safety is a crucial factor, which must be guaranteed for the entire life cycle of project management. For example, in the field of cross-organizational collaborative decision-making, there is a great deal of private information that companies are reluctant to leak, even when such information is needed for collaborative data analysis. This issue is emphasized on the one hand by the lack of adequate mechanisms for protecting privacy in cross-organizational collaborative decision-making processes and on the other by the ever-increasing use of big data \cite{zhu2017privacy}. Similarly, the same issues described above apply to workflow management, which is crucial for improving business productivity. Indeed, many workflow systems go outside the organizational boundaries and often require organizations to interact with each other. Each organization has its own private business processes and can operate autonomously, but at some point, all the organizations involved need to be synchronized to complete certain tasks. It is easy to imagine how such organizations are unwilling to share business details with others \cite{liu2020privacy}. Another non-negligible problem in this context is that while some organizations may be allied for a project, the same organizations may be competitors for other projects \cite{warnier2010intellectual}.
Furthermore, ever-increasing security issues are emerging regarding cross-organizational cooperation in ubiquitous computing environments, mainly due to the interoperability problems deriving from the different security mechanisms and policies put in place by each organization \cite{hilia2011cross}. Very often, the implementation of cross-organizational business processes requires systems that allow federated identity management. Indeed, in such processes, there are administrative domains of different partner organizations that need to interact with each other, and all this, in some way, requires that the partners trust each other \cite{thurm2008automated}.

Blockchain \cite{Blockchain} is a data storage technology that originated from Bitcoin, a peer-to-peer cryptocurrency \cite{nakamoto2012bitcoin} that realizes block synchronization through peer-to-peer transmission technology and consensus algorithm, ensuring the data consistency of each member node in the network. The tampering resistance of the data registered in the Blockchain network against external attacks has been proven to be efficient \cite{liu2020fabric}. The data state is read or changed through transactions assembled and packaged into blocks under a specific structure in a Blockchain network. Each block keeps the previous block's hash value, so if any block's hash value is changed, the entire chain will be invalidated. Depending on the level of trust between nodes, Blockchains can be divided into a \emph{public chain}, \emph{consortium chain}, and \emph{private chain}. The nodes of a private chain all belong to the same organization and are fully trusted. The nodes of a consortium chain belong to different organizations that trust each other, and lastly, all nodes of a public chain do not trust each other. Hyperledger Fabric is an open-source Blockchain platform that can be used to implement consortium chain networks \cite{10.1145/3190508.3190538}, and realizes all characteristics of Blockchain, including decentralization, irreversibility, consensus, identity authentication, smart contract, and others. Compared with other Blockchain platforms, Hyperledger Fabric provides higher throughput, a more effective consensus mechanism, a channel isolation mechanism, a multi-chain mechanism, and flexible expansion capability.
Several studies use Blockchain as the underlying data platform to solve information-sharing problems among project participating organizations \cite{p3,p2,p1}. In particular, Liao et al. have proposed a Blockchain-based cross-organizational integrated platform, called \emph{BCOIP} \cite{liao2019Blockchain}, which enables to issue and redeem of reward points. Lu et al. use Blockchain technology to store users' access control lists. In this way, thanks to its tamper-proof and decentralized features, Blockchain technology allows the creation of cross-organizational authentication systems where organizations can share data and resources between them \cite{lu2018privacy}. Again, Fridgen et al. show how Blockchain can be a viable solution to achieve secure cross-organizational workflow management \cite{fridgen2018cross}. In particular, Blockchain in business process management allows improving the auditability and automation of manual processes through a decentralized system. Furthermore, it is essential to underline that the development and deployment of Blockchain-based systems for cross-organizational workflows management cannot ignore the legal regulations regarding data processing, such as the \emph{General Data Protection Regulation (GDPR)} in force in Europe \cite{guggenmos2020develop}. However, most of the studies proposed in the literature are based on domain-specific implementations and do not provide a generic management tool that project managers can reuse.

In this paper, we propose a general-purpose project management system, referred to as \emph{Fabric-GC}, realized using Blockchain as a data-sharing platform. More precisely, the proposed system uses the Gantt chart model to manage the entire project allocation and execution progress, besides visually providing such relevant information to project managers. Again, \emph{Fabric-GC} applies Blockchain technology so that project data can be shared safely and efficiently among multiple organizations, facilitating cross-organizational project collaboration. In detail, \emph{Fabric-GC}, which represents the first Gantt chart management system for cross-organizational project management, is based on hyperledger fabric. The consortium chain is selected as the underlying storage model for the system proposed.
The main contributions of this article are as follows:

1) Blockchain and Gantt chart are the building blocks of \emph{Fabric-GC}. The proposed solution enables the migration of the traditional Gantt chart model from a centralized to a distributed architecture to provide visual expression. Besides, the Blockchain is also tackled to deal with the secure storage and sharing of data, where smart contracts define the structure and operation of data in a project.

2) The proposed solution referred to as \emph{Fabric-GC} aims at dividing the entire project into multiple chunks of small tasks. The project manager defines the project plan and assigns such chunks to different organizations in task schedules; then, it uses smart contracts to specify the read and write operations on the project plan. The proposed solution effectively improves the flexibility of project cooperation and guarantees versatile project management.

3) The proposed solution enables the visualization of task schedules as a Gantt chart, besides providing a progress feedback mechanism that assists project managers in grasping the project completion status and making real-time adjustments to the project plan.

4) Experimental results show that Fabric-GC has stable performance and high production efficiency under different consensus mechanisms.

The remaining of this article is organized as follows. Section \ref{sec:relatedWork} introduces the work related to this proposed research, Section \ref{sec:preliminaries} presents some preliminary concepts, including the data storage mechanism of Hyperledger fabric and structure of the Gantt chart. Section \ref{sec:system} introduces the system architecture, data structure, smart contract design, and workflow. Section \ref{sec:experiment} discusses the operation steps of the Fabric-GC system and shows the system's stability under different consensus mechanisms through comparative experiments. Finally, Section \ref{sec:conclusion} summarizes our contributions and brings items as future work.

\section{Related Work}
\label{sec:relatedWork}

Widely speaking, Blockchain technology enables the realization of decentralized, immutable, and incorruptible public ledgers \cite{p4}. Due to its ability to create smart contracts, Blockchain is perfectly suitable for project management, which phases include project creation, project allocation, project execution, and project acceptance. As known, the entire project cycle requires information sharing and oversight from multiple parties. In this context, the ability to access electronic data securely and efficiently enhances the ability to perform quality assurance-type projects. Therefore, the applicability of Blockchain in project management has been investigated by many researchers, as shown in Table \ref{tab:related}, which summarizes these studies.

\begin{table}[htbp]
	\caption{Comparison with related work.}
	\begin{center}
		\label{tab:related}
		\resizebox{\textwidth}{15mm}{
			\begin{tabular}{c|c|c|c|c}
				\toprule
				\textbf{Research} & \textbf{Application} & \textbf{Model} & \textbf{Generality} & \textbf{Visualization} \\
				\midrule
				\cite{YANG2020103276}, \cite{8792582} & Construction Engineering & N & N & N \\
				\cite{8623571}, \cite{Meng2020} & Scientific Research Project Management & N & N & N \\
				\cite{HELO2020101909} & Supply Chain & N & N & Y \\
				\cite{LIU2020101897} & industrial Production & N & N & N \\
				\cite{ljmu13015}, \cite{aaaaaa} & Government Project Management & N & N & N \\
				This paper & General Project Management System & Y & Y & Y \\
				\bottomrule
		\end{tabular}}
	\end{center}
\end{table}

To address the issues of poor communication, weak file sharing privacy, and low-quality submission efficiency in projects construction, Yang et al. \cite{YANG2020103276} analyzed the business processes of the public and private Blockchain in the construction industry and presented the challenges faced by the construction industry after applying Blockchain, aimed at improving the efficiency and productivity of construction projects. In addition, Hargaden et al. \cite{8792582} proposed to apply Blockchain to sizeable structural engineering projects. They concluded that incorporating Blockchain improves efficiency, trust, transparency, and regulation in the construction industry effectively. However, the above works proposed in the literature do not introduce a specific system model to address the multi-party project management problem. 
Scientific and engineering project works also need strict regulation and monitoring to reduce human communication and supervision costs \cite{p7}. Bai et al. \cite{8623571} proposed a Blockchain-based \emph{scientific research project management system (SRPMS)} and analyzed the five functional modules of the proposed model. Meng et al. \cite{Meng2020} used consortium Blockchain and \emph{IPFS (InterPlanetary File System)} technology to realize a reliable and efficient scientific research project management system that overcomes the limitations on the breach of contract and confidentiality in project management, also reducing the time and labor cost for the project implementation. Helo et al. \cite{HELO2020101909} applied Blockchain in the supply chain to solve the delivery problem of multi-supplier participation by ensuring real-time tracking, control of data, and real-time visibility of all the processes in the project production process under the control of a project manager. Liu et al. \cite{LIU2020101897} used Blockchain to manage the life cycle of products in the industrial production process, enabling the coordinating production information across departments and partners, quickly and accurately tracking the production and sales process, improving interoperability and collaboration among stakeholders in the product chain. To cope with several government-supported projects, Lee et al. \cite{ljmu13015} proposed a generic project sharing platform that achieves project information sharing while ensuring the platform's anti-forgery with the help of \emph{POA (Proof-of-Authority)} consensus algorithm. Lastly, Green \cite{aaaaaa} showed that the adoption of Blockchain in the digital management of government projects could significantly improve workload and productivity, besides improving the strategic decision-making of the government.

The abovementioned issues show that Blockchain technology can be applied to the project management process to improve many aspects such as collaboration capability, information security, and real-time tracking functions of project implementation in a multi-organizational cooperation mode, to enhance the project completion efficiency. 
Several studies indicate that decomposing large projects into multiple small task schedules and sequencing the execution of task sets in a time series can achieve rational resource planning, besides saving time and labor costs \cite{6596335, doi:10.1287/ited.2016.0168, pr7120917, 10.1007/978-3-319-09339-0_24, JIA2007313, 7507546}. 
Similarly, Blockchain is applied in ensuring secure data storage in areas such as online education, finance, Internet of Thing (IoT), healthcare, and Vehicular ad-hoc network (VANET) \cite{liu2020blockchain, li2022moocschain, cui2019efficient, li2020fabric, li2021privacy, li2020privacy, liu2021behavior, han2021blockchain, li2021fabric, zhijie2022blockchain, han2020traceable, cui2020arfv, li2022blockchain, sun2022blockchain}.
However, the current strategies have not saved project costs from the details of rational planning projects, nor providing sufficient simulation experiments to demonstrate performance sustainable performance under large-scale and multi-harmonic tasks.
Thus, in this paper, we proposed a generic distributed management tool for project managers to adequately handle the management and coordination of decentralised, complex or large projects.

\section{Preliminaries}
\label{sec:preliminaries}

This section presents the background and related methods for the system's design and implementation to give further details of the proposed system. The notation used in this article is outlined in Table \ref{tab:notations}.

\begin{table}[h]
	\centering
	\caption{The descriptions of notations.}
	\label{tab:notations}
	\begin{tabular}{cc}
		\toprule
		\textbf{Notations} & \textbf{Description} \\
		\midrule
		$u_i$ & External user $i$ \\
		$p$ & Represent a project \\
		$t_{j}$ & $j$-th task scheduling of $p$ \\
		$P$ & Project list \\
		$T^n$ & A set of $n$ tasks \\
		$bPT$ & Start time of the project \\
		$ePT$ & End time of the project \\
		$bT$ & Start time of the task \\
		$eT$ & End time of the task \\
		$cT$ & Completed time of the task \\
		$uN$ & User name \\
		$tN$ & Task name \\
		$pN$ & Project name \\
		$PI$ & Index of user and projects \\
		\bottomrule
	\end{tabular}
\end{table}

\subsection{Hyperledger Fabric}
Hyperledger Fabric is used to build enterprise-level consortium Blockchain and realize data sharing among multiple organizations to collaborate to form Blockchain networks. As an open-source project, Hyperledger Fabric has been started by the Linux Foundation and maintained by several corporate organizations. Basically, Hyperledger Fabric is characterized by a modular design concept. It has a sophisticated tiered policy structure, where each fabric component is extensible and mainly includes identity authentication, consensus module, intelligent contract, data storage. Each component is a container, so it is high the flexibility to build the fabric network. The entire system runs in the docker container. The container separates the running environment from the hardware environment as a sandbox environment to achieve total data confidentiality and security. 
The protocol used for the secure channel is TLS (Transport Layer Security). 
TLS/SSL is a specification for an encrypted channel that uses symmetric encryption, public-private key asymmetric encryption.
Finally, all nodes in the fabric network need authentication and authorization. These requirements enable the meeting of the characteristics of mutual trust among members of the consortium Blockchain.

\subsubsection{Main Components}
There are three most important types of nodes in Hyperledger Fabric: CA, Orderer, and Peer.

\textbf{CA:} In fabric networks, the identity certificate is required for communication. Without loss of generality, we assume that external users intend to communicate with one of the nodes. 
The CA acts as a trusted entity and holds the public keys of all users, but the algorithm for generating public and private key pairs for user registration is executed locally. 
In that case, it needs to be registered by the administrator at the CA node to generate a unique digital certificate and key for data transmission.

\textbf{Orderer:} It is mainly responsible for data consensus, and all validated transactions are submitted to the Orderer node for sorting. Next, the Orderer node packages the transactions into blocks according to the predefined rules (block out time, block size, the maximum number of transactions, etc.) and then sends them to the Peer node. A consensus algorithm maintains the consistency of data, in which consensus mechanisms provided by the fabric are \emph{Solo}, \emph{Kafka}, and \emph{Raft}.

\textbf{Peer:} It is a data storage node, either for Blockchain state or block data. In addition, it has the function of validating transactions by simulating the execution of chaincodes to verify the legitimacy of transactions, i.e., endorsement. Only legitimate transactions are submitted to the Orderer node waiting to be packaged into blocks. Lastly, their modifications on the state are written to the Blockchain.

\subsubsection{Chaincode}
Chaincode is the smart contract of fabric and is implemented mainly using the Go programming language. It is the interface of Blockchain to the external environment. By calling the methods defined by chaincode, the external environment can execute operations such as data storage, indexing, or modification to the Blockchain, which functionally is similar to SQL language in relational database \cite {10.1007/978-3-030-20948-3_17}. Developers writing different chaincode programs can achieve different application functions.

\subsubsection{Ledger}
\begin{figure}[ht]
	\centering
	\includegraphics[width=\linewidth]{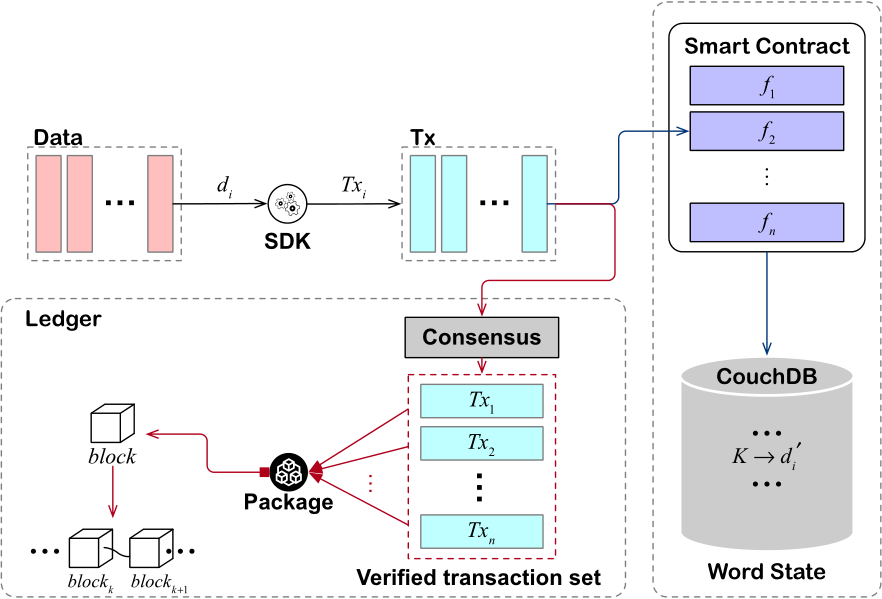}
	\caption{Structure of the ledger in Hyperledger Fabric.}
	\label{fig:ledger}
\end{figure}

There are two types of data on fabric, the world state and block. As shown in Fig~\ref{fig:ledger}, external data $d_i$ is packaged as a transaction operation $Tx_i$ by calling the SDK \cite{fabric-sdk-go, fabric-sdk-node, fabric-sdk-java, fabric-sdk-py}. Then, the transaction writes $d_i$ to the world state by calling the method $f_i$ in the smart contract and is stored to the state database in the form of $<k-v>$.

The legal $Tx_i$ will be submitted to the Orderer node, waiting to be packaged into blocks and stored permanently by the Peer node.

\subsection{Gantt Chart}
Gantt chart is a management tool for planning and project arrangement proposed by Henry Gantt \cite{6596335}, widely used in many fields, such as educational activities, software development \cite{7507546}, technology transfer \cite{pr7120917}, production plant scheduling \cite{JIA2007313}, and several others. Gantt chart shows graphically the project plan, which can be handy to track the task scheduling in each period.

The horizontal axis of a Gantt chart represents time, and the vertical axis represents task scheduling. For a project $p$, it can be divided into $n$ small task schedules based on time, resources, manpower, etc., and $p=\{t_1,t_2,\cdots,t_n\}$. If these tasks are scheduled to be executed only in time order, the total execution time of a project can be characterized as follows.

\begin{equation}
	\begin{aligned}
		\Delta t &= p.ePT-p.bPT \\
		&=(t_1.eT-t_1.bT)+(t_2.eT-t_2.bT)+\\&\cdots+(t_n.eT-t_n.bT) \\
		&=\sum_{i=1}^{n}{t_i.eT-t_i.bT}
	\end{aligned}
\end{equation}

In the case we analyze the key execution order $\{t_{s1},t_{s2},\cdots,t_{sk}\}$ in the task set \cite{pr7120917}, where $s1\leq s2\leq \cdots\leq sk$ and $\{s1,s2,\cdots,sk\}\subset[1,n]$, then, the overall project execution time is given as follows.

\begin{equation}
	\begin{aligned}
		\Delta t_{theory} &= (t_{s1}.eT-t_{s1}.bT)+(t_{s2}.eT-t_{s2}.bT)+\\&\cdots+(t_{sk}.eT-t_{sk}.bT) \\
		&= \sum_{i=1}^{k}{t_{si}.eT-t_{si}.bT},\quad 1\leq k\leq n
	\end{aligned}
\end{equation}

Gantt chart can visualize the execution relationship between each task schedule. Due to such, project managers utilize the Gantt chart to plan and adjust the project execution. In this way, they can potentially more accessible estimate the project cost, evaluate the project deadline, and achieve or approach the theoretical time cost $\Delta t_{theory}$.

\begin{figure}[ht]
	\centering
	\includegraphics[width=\linewidth]{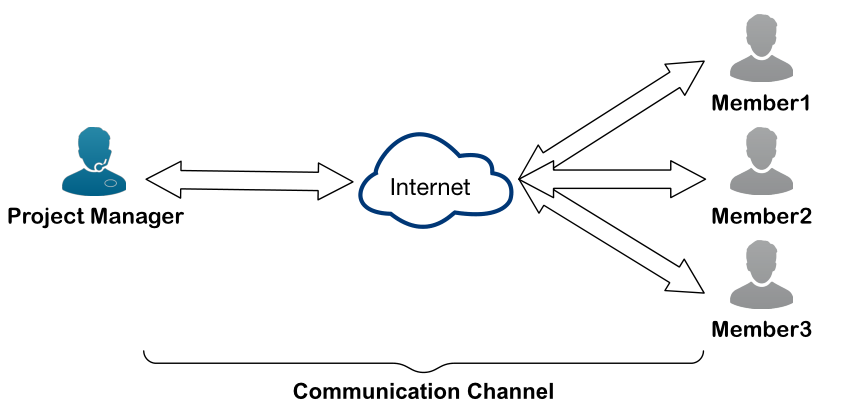}
	\caption{The centralization mode of information interaction.}
	\label{fig:center}
\end{figure}

Although the Gantt chart achieves excellent performance in project planning, most of the current Gantt chart systems in the market utilize a centralized model, as in Fig~\ref{fig:center}. Therefore, when multiple organizations are involved in a project, problems such as lagging news and untimely feedback inevitably occur. The unsynchronized information may cause management moil and bring severe negative impact to projects.

\section{System Architecture and Design} 
\label{sec:system}
It is introduced in this section the architecture of Fabric-GC. The overall design architecture is discussed first, followed by the data structure defined in Fabric-GC. Next, the design method of the smart contract, and lastly, the workflow of Fabric-GC.

\subsection{System Architecture}

\begin{figure}[htbp]
	\centering
	\includegraphics[width=\linewidth]{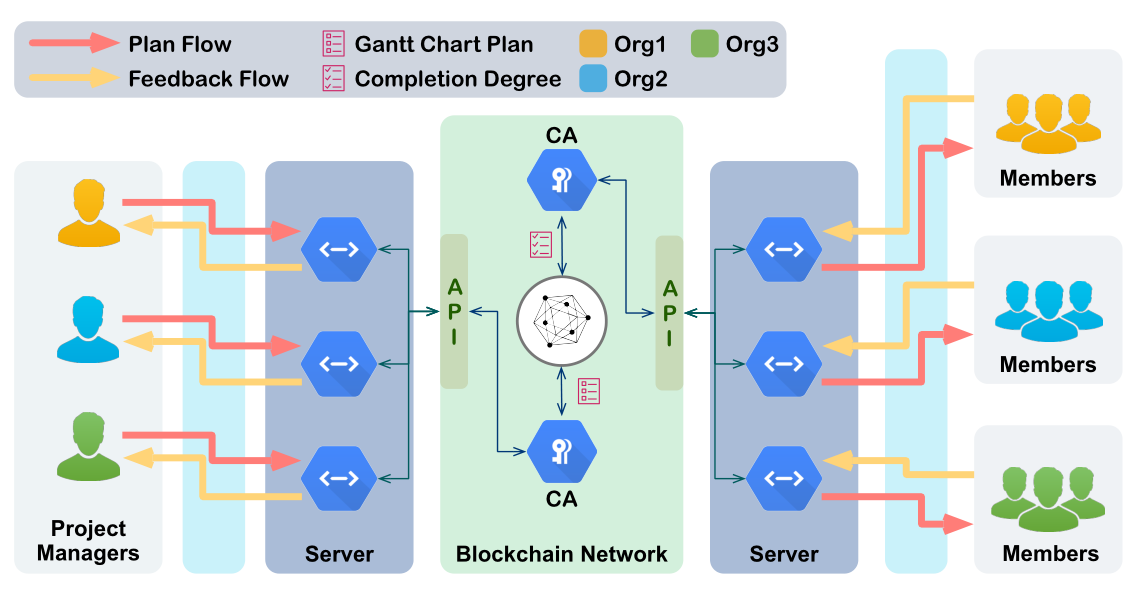}
	\caption{Architecture of fabric-GC.}
	\label{fig:System Architecture}
\end{figure}

The Blockchain-based Gantt chart system proposed in this work consists of three parts: \emph{consortium Blockchain}, \emph{server layer}, and \emph{user layer}. The functions of Fabric-GC include permission to participants from different organizations to join the system, project plan sharing in the form of Gantt charts, and feedback from project members on the project execution progress.

\textbf{Consortium Blockchain:} a distributed network composed of nodes representing different organizations for global data synchronization and storage. The nodes in the consortium are mutually trusted. More precisely, they realize identity verification through digital certificates to ensure the security and integrity of data in the system. The smart contract running on it regulates the various steps in project management and stores the project data in Ledger for permanent storage.

\textbf{Server layer:} It contains servers maintained by each organization, interacts with a Blockchain system and smart contracts, and provides an endpoint interface to project members of the same organization. Project data is removed from the Blockchain and converted into meaningful project plans at that layer and visualized as Gantt charts provided to the project manager.

\textbf{User layer:} It contains all project members and includes project managers and project participants. Project managers are responsible for the planning, procurement, and execution of a project, in any undertaking that has a defined scope. Project participants follow the project arrangement and feedback on the project progress. However, in Fabric-GC, the user identity is not distinguished, and thus, the design eliminates the organizational differences and realizes the conversion of logical identity.

As shown in Fig~\ref{fig:System Architecture}, there are two different data streams in Fabric-GC: \emph{Plan Flow} and \emph{Feedback Flow}. According to the existing project resources, the project manager seeks to achieve defined goals by using plans, schedules the project execution, and then draws the Gantt chart. This chart is then submitted to the Blockchain system through Plan Flow, so other project participants can obtain specific project plans from the Blockchain and complete the assigned tasks of the project according to the project arrangement and schedule. Any modification in the project execution processing will notify project members in time. When the project members conduct the project, they submit the completed progress through Feedback Flow, also feedback it to the project manager through the Blockchain system. Upon receiving such relevant updated information, the project manager makes appropriate adjustments to the plan according to the progress. In this way, closed-loop data exchange is formed to realize the dynamic management of projects across organizations.

\subsection{Data Structure}

\begin{figure}[htbp]
	\centering
	\includegraphics[width=\textwidth]{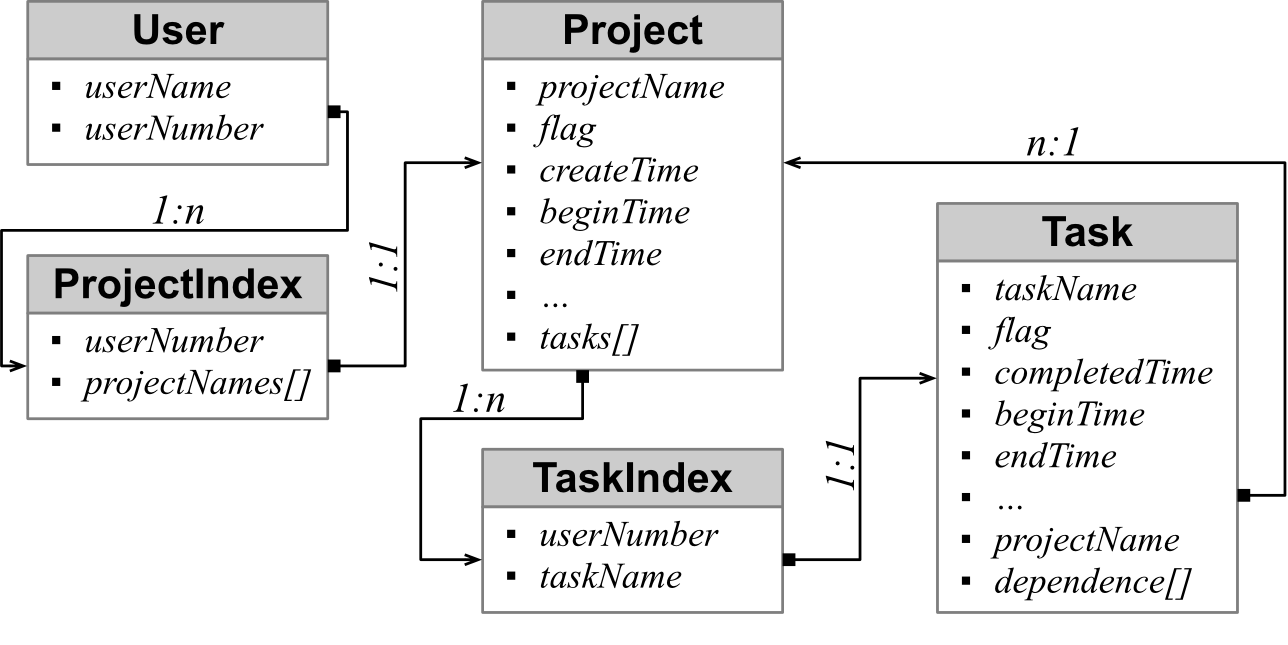}
	\caption{Relationship between data structures.}
	\label{fig:Data Struct}
\end{figure}

The world state of Blockchain is similar to table data in a relational database. The data structure of the state is analogous to the table structure \cite{10.1007/978-3-030-20948-3_17}. Five data structures are defined in this article, listed as \emph{Project}, \emph{Task}, \emph{ProjectIndex}, \emph{TaskIndex}, and \emph{User}. As shown in Fig~\ref{fig:Data Struct}, there is an index relationship between the data structures, and depicted in Eq. (\ref{eq:index}). We remark that defining the data structures in this way facilitates uniform data access operations and reduces the system's complexity.

\begin{equation}
	\label{eq:index}
	\begin{cases}
		User\rightarrow ProjectIndex &= 1:n \\
		ProjectIndex \rightarrow Project &= 1:1 \\
		Project \rightarrow TaskIndex &= 1:n \\
		TaskIndex \rightarrow Task &= 1:1 \\
		Task \rightarrow Project &= n:1
	\end{cases}
\end{equation}

\subsubsection{User}
This structure defines a project member in the Blockchain system, and only project members added to the Blockchain can be authorized to participate in relevant projects. The User structure contains two fields, $u_i=\{userName,\ userNumber\}$, where $userName$ represents the user name used by all participants to identify member $i$, and $userNumber$ denotes the unique identification of member $i$ when data is stored.
Also, to enhance the uniqueness and randomness of the $userNumber$, we introduce a timestamp so that the $userNumber$ can be calculated as calculated by Eq. (\ref{eq:e1}).

\begin{equation}
	\label{eq:e1}
	userNumber=Hex[Hash(pubKey_i, Timestamp)]
\end{equation}

\subsubsection{Project}
This structure holds the properties of the project, obtained from $pN$. The essential attributes of this structure are as follows.

i) $flag$: records the status of the current project. $processing$ indicates that the project is in progress, and $done$ indicates that the project is completed. The default value is $processing$.

\begin{equation}
	\label{eq:flag}
	flag=\begin{cases}
		processing \\
		done
	\end{cases}
\end{equation}

ii) $beginTime(bPT)$: Project start time used to limit the left interval of task scheduling.

iii) $endTime(ePT)$: Project end time, used to limit the right range of task scheduling.

iv) $tasks$: The collection of all tasks for this project.

\subsubsection{Task}
This structure represents the data structure of task scheduling, which stores the attributes of task scheduling and is indexed by $tN$. The task set of $p$ is expressed as $T^{n}$, then $p$ expressed according to Eq. (\ref{eq:e2}).

\begin{equation}
	\label{eq:e2}
	p=T^{n}=\{t_{1},t_{2},\dots,t_{n}\}
\end{equation}

Additionally, the essential attributes of the task structure are as follows.

i) $flag$: Record the status of this task schedule. From Eq. (\ref{eq:e2}), $p.flag=done$ is equivalent to

\begin{equation}
	\forall t_{j}\in T^{n},\ t_{j}.flag=done
\end{equation}

ii) $beginTime(bT)$: The start time of the task schedule. The value range of this attribute is $bT\in [bPT,ePT)$.

iii) $endTime(eT)$: The end time of the task schedule. The value range of this attribute is $eT\in (bPT,ePT]$ and $bT<eT$. At the same time, it must satisfy the following equation:

\begin{equation}
	\begin{aligned}
		t_{n}.eT-t_{1}.bT&\leq p.ePT-p.bPT \\
		t&\in T^{n}
	\end{aligned}
\end{equation}

iv) $completedTime(cT)$: The completion degree of the current task schedule. It is noted that, if and only if $cT=eT$, $flag=done$ holds.

v) $dependence$: In actual task scheduling, the start of a task may require completing other tasks, so this property saves the set of dependent tasks for the current task.

\subsubsection{ProjectIndex}
Fabric-GC allows multiple projects to co-exist in the system. A member can participate in multiple projects, so the structure defines two properties: $userNumber$ and $projectNames$. The former identifies a member, and the latter records the name of each project the member participates in.

\subsubsection{TaskIndex}
The structure is saved in the $tasks$ of the Project. Two attributes are defined, where $userNumber$ represents the member responsible for scheduling the task, and $taskName$ records the name of the task schedule and can index the entire task scheduling data.

\subsection{Smart Contract Design}
The contract part mainly defines data access operations and relies on the conventional MVC (Model-view-controller) software design pattern \cite{Voorhees2020}. This part avoids using excessive business processing logic to reduce functional redundancy and improve the system's scalability. In this research, the following methods are defined to access the data corresponding to members, projects, and task scheduling, respectively.

\subsubsection{Member Data Access}
There are two ways to register members in the contract: $createUser()$ and $queryUser()$. Again, to ensure the uniqueness of the identity of participants belonging to multiple organizations, a member is stored in the state database of the Blockchain. Meanwhile, before creating and indexing a member, the system checks whether the currently created member has already been stored in the Blockchain system.

There are four methods related to project data in the contract. listed as: 
\begin{enumerate}
	\item $createProject()$ writes the data of the project to the state database. Before creating the project, we must check whether the project already exists. After the new project is successfully created, a ProjectIndex should be established between the creator and the project. 
	\item $queryProIndex()$ takes the member as the input to obtain the project name set participated by the member. 
	\item $queryProject()$ indexes the data of the project through the project name. 
	\item $changeProject()$ is used to index the project data and modify the data of the current created project to realize the flexibility of data access.
\end{enumerate}

\subsubsection{Task Data Access}
There are three contract methods related to task scheduling data, listed as:

\begin{enumerate}
	\item $assignTask()$ allocates task scheduling for the specified project. After successful creation, the TaskIndex needs to be saved to the $tasks$ of the project. Then, the ProjectIndex is established for the members responsible for the task. 
	\item $queryTask()$ method obtains the specific task scheduling data through the task name. 
	\item $changeTask()$ can modify the specified task scheduling information, as shown in Algorithm \ref{al:a1}. The modification of scheduling can be divided into two categories: when the value of $target$ is "changeInfo", it indicates that only the data of the current task scheduling needs to be modified. When the value of $target$ is "changeManager", the current project leader needs to be modified. At this point, the information of the task scheduling and the project attribute and project index related to the task must be modified.
\end{enumerate}

\begin{algorithm}[ht]
	\caption{changeTask}
	\label{al:a1}
	\KwIn{$tN$, $target$, taskData}
	\KwOut{}
	
	\If{$target$ == "changeInfo"}{
		DelState($tN$);\\
		PutState($tN$,taskData);
		
		\KwRet{};
	}\Else{
		\If{$target$ == "changeManager"}{
			oldTData = GetState($tN$);\\
			project = GetState(oldData.projectName);\\
			\While{t in project.tasks}{
				\If{t.taskName == tN}{
					t.userNumber = taskData.manager;
				}
			}
			PutState(oldData.projectName,project);\\
			createProjectIndex(taskData.manager, oldData.projectName);\\
			PutState($tN$,taskData);
			
			\KwRet{};
		}
	}
\end{algorithm}

\subsection{Workflow}
The system operation is divided into three logical parts to enable the Blockchain-based Gantt chart (i.e., Fabric-GC). Aimed to realize the cross-organizational project management function, it consists of participant login, project creation, and task scheduling allocation. In the following, we provide the details of the three parts abovementioned.
%

The complete participant login processing can be divided into three parts: administrator registration, project member registration, and project member login. This process can realize the storage of personnel information among different organizations on the Blockchain and solve relatively weak information exchange among members of different organizations.
In detail, each organization has a Fabric CA node used to store each member's ID, private key, certificate, and other information. Before registering a member, we need to register an administrator user to connect to the Fabric CA node. 
The steps to carry out this operation are as follows.

\begin{enumerate}
	\item The server creates a $Wallet$ locally, 
	\item sets the administrator's name and password and sends it to the fabric CA node of the organization through the SDK, 
	\item Fabric CA node generates unique $signID$, $privKey_{admin}$, $pubKey_{admin}$ and certificate for the administrator, 
	\item sends these to the server and saves them in the created $Wallet$ as permanent storage. 
\end{enumerate}

\begin{algorithm}[ht]
	\caption{Register $u_i$ in the system.}
	\label{al:a2}
	\KwIn{$u_i$.userName, $u_i$.Org}
	\KwOut{}
	
	$//$ Register the $u_i$.\\
	Send {$u_i$.userName, $u_i$.Org} to the server;\\
	Call the SDK specified by $u_i$.Org;\\
	\If{!isExists($u_i$.userName)}{
		ca $\leftarrow$ connect(\{admin, adminpw\});\\
		CA generates \{$signID$, $pubKey_i$, $privKey_i$, $certificate$\};\\
		CA sends \{$signID$, $pubKey_i$, $privKey_i$, $certificate$\} to the server;\\
		The server save them to the wallet;\\
		$u_i$.userNumber $\leftarrow$ Hex[md5($pubKey_i$)];\\
		SDK.createUser($u_i$.userName, $u_i$.userNumber);\\
		\KwRet{};\\
	}
	\Else{
		\KwRet{'An identity for $u_i$ already exists in the wallet.'};
	}
\end{algorithm}

Notably, Administrators do not need to be stored on the Blockchain to distinguish different organization members and reduce unnecessary data conflicts.
Algorithms \ref{al:a2} and \ref{al:a3} show how to register and log $u_i$ into Fabric-GC. When the members are registered, $u_i$ needs to provide the $uN$ and the organization's name to which it belongs. 
Then, according to the organization name, the server selects the corresponding organization's SDK to call. 
Before a member is registered, it is necessary to check whether it has been registered and whether $uN$ already exists in the CA node. If the member is not registered, the server logs into the Fabric CA node through the administrator account and password, then the CA node registers $u_i$ and generates the field $\{signID, pubkey_i, privKey_i, certificate\}$, and then sent to the server. Finally, the server calculates the $usernumber$ by using the Eq. (\ref{eq:e1}) and executes the $createUser()$ method of the smart contract to invoke the member $u_i$ into the Blockchain state database, completing the registration process of the project members.

\begin{algorithm}[ht]
	\caption{Log in $u_i$ in the system.}
	\label{al:a3}
	\KwIn{$u_i$.userName, $u_i$.Org}
	\KwOut{}
	
	$//$ Log in $u_i$.\\
	Send {$u_i$.userName, $u_i$.Org} to the server;\\
	Call the SDK specified by $u_i$.Org;\\
	\If{isExists($u_i$.userName)}{
		Get $\{privKey_i,pubKey_i\}$ from the wallet by $\{u_i.userName, admin\}$;\\
		Connect to fabric network by $\{privKey_i,pubKey_i\}$.\\
		\{$u_i$.userName,$u_i$.userNumber\} $\leftarrow$ SDK.queryUser($u_i$.userName);\\
		\KwRet{};
	}
	\Else{
		\KwRet{'An identity for $u_i$ does not exists in the wallet'};
	}
\end{algorithm}

During the member login process, the $uN$ and the organization name are also provided by $u_i$. The server will first determine whether the member exists. If the result provided by the query locates such a member, $u_i$ can connect to the Blockchain network through the public key $pubKey_i$ and private key $privKey_i$, and then query the member data by calling the $queryUser()$ method of the smart contract. If successful, the data will be sent to the server, and the server responds with the login results to $u_i$.

\subsubsection{Project Creation}

\begin{figure}[ht]
	\centering
	\includegraphics[width=\textwidth]{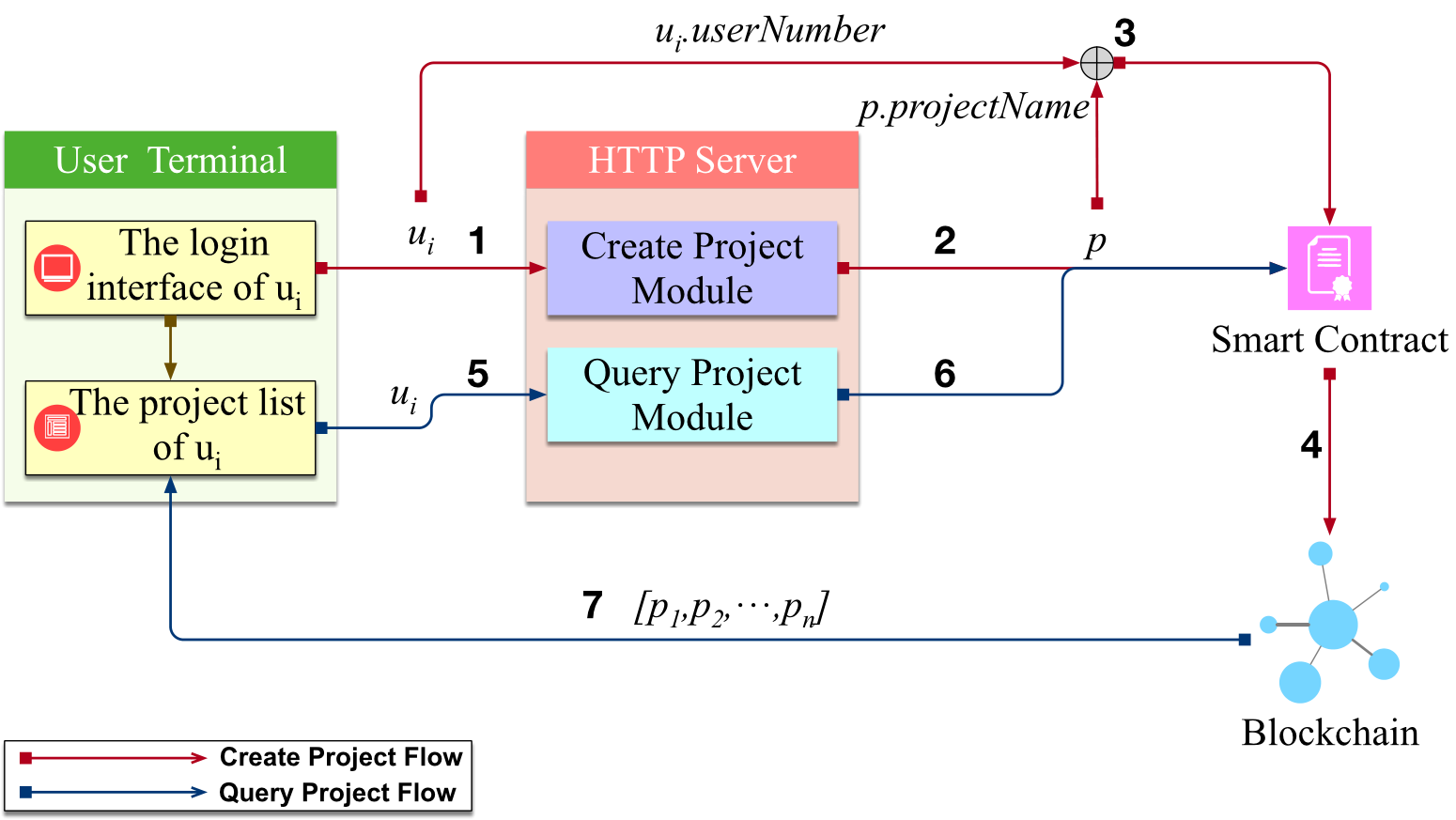}
	\caption{The description of project creation process.}
	\label{fig:Create Project}
\end{figure}

Solely when the project is created in the Fabric-GC, and relevant attributes of the project are specified, the project manager can assign task scheduling for the project and draw a Gantt chart. The project creation process of the system is outlined in Fig~\ref{fig:Create Project}.

\textbf{Step 1:} Member $u_i$ accesses the server and sends $\{userName_i, Org_i\}$ to the server's 'process\_login' module to requests system login. After the server responds to the successful login, $u_i$ sends the project creation request to the "create\_project" module and submits the project-specific attribute values.

\textbf{Step 2:} The server generates Project data structure $p$. Next, it sets $p.tasks=null$ and checks whether the value of $p.flag$ is $processing$. If the data format meets the requirements, the chaincode $createProject()$ is called through the SDK specified by $Org_i$, and the parameter \{$userNumber_i$, $JSON(p)$\} is passed in. Besides, if isExists($p$) is false, it means that the project has not been created before and PutState(\{$p.projectName$, $p$\}) is called to create the project.

\textbf{Step 3:} After the project is created successfully, the index $PI$ of $userNumber_i$ and $p.projectName$ is created by $createProIndex()$.

\textbf{Step 4:} Project $p$ and index $PI$ are stored in the state database of Blockchain.

\textbf{Step 5:} Member $u_i$ sends a request to the server to query the list of projects that participates in.

\textbf{Step 6:} After receiving the request, the server executes the "query\_project" module and indexes the data based on Eq. (\ref{eq:e3}).

\begin{equation}
	\label{eq:e3}
	\begin{aligned}
		u_i &\xlongrightarrow{uN_i} \{u_i,[pN_{1},pN_{2},\dots,pN_{n}]\} \\
		&\xlongrightarrow{pN_{j}} [p_{1},p_{2},\dots,p_{n}], \quad j=1,2,\dots,n
	\end{aligned}
\end{equation}

The chaincode $queryProIndex()$ is called by SDK specified by $Org_i$, and the list of project names $PN= [pN_{1},pN_{2},\dots,pN_{n}]$ is obtained. After the query is successful, the server iteration $PN$, and the chaincode queryProject ($queryProject(pN_{j})$), $j=1,2,\dots,n$ is called to obtain all the list of projects $P=[p_{1},p_{2},\dots,p_{n}]$.

\textbf{Step 7:} The project list $P$ is sent to the client terminal for visualization by member $u_i$.

\subsubsection{Task Schedule Allocation}
\begin{figure}[ht]
	\centering
	\includegraphics[width=\linewidth]{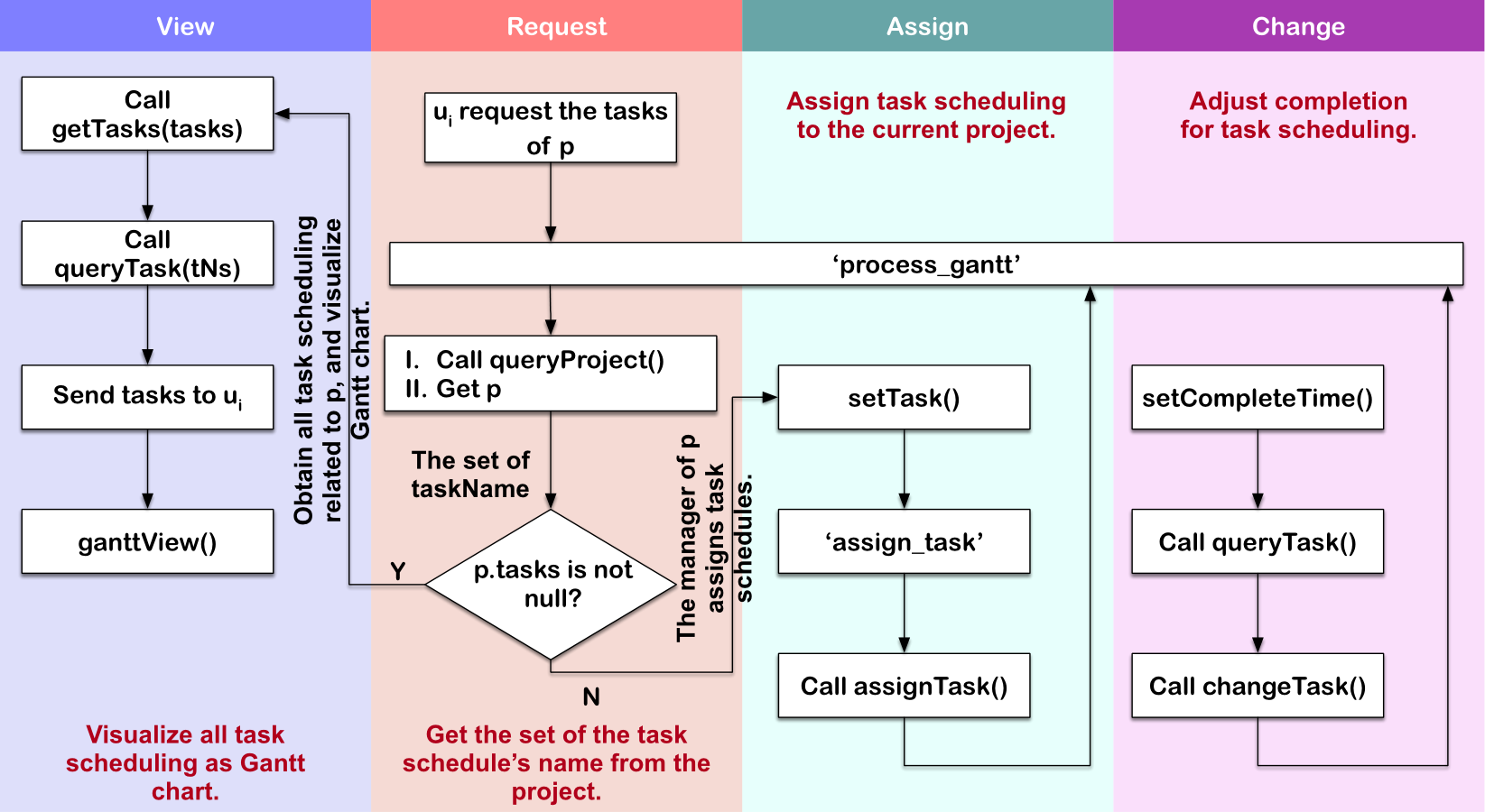}
	\caption{The flow of task schedule allocation process.}
	\label{fig:assignTask}
\end{figure}

The task schedule allocation is the most critical part of the Fabric-GC proposed system. The main characteristic of this part is that it enables to achieve sharing and dynamicity. Besides, the underlying Blockchain also ensures the integrity and synchronization of data in the project management. As Fig~\ref{fig:assignTask} shows, task schedule allocation is divided into four modules: data request, task scheduling allocation, completion modification, and Gantt chart visualization. A detailed description of these four modules is provided next.

\textbf{Request:} As soon as the member $u_i$ retrieves the list of projects it participates in, a project $p$ is selected and requests next the tasks from the server. Afterward, the server runs the "process\_gantt" module by calling the $queryProject()$ method of the chaincode through the SDK to obtain the task name set $p.tasks$ of $p$. If $p.tasks$ is $null$, it means that the current project $p$ has no tasks assigned yet, and the project leader needs to make a reasonable task scheduling assignment.

\begin{algorithm}[ht]
	\caption{Assign task scheduling to $p$}
	\label{al:a4}
	\KwIn{$tN$, manager, $bT$, $eT$, $flag$, info, dependence}
	\KwOut{}
	
	The server runs the 'assign\_task' module;\\
	$cT$ = null;	$//$ The task does not begin.\\
	$pN$ = $p$.projectName;\\
	$T'$ $\leftarrow$ Call SDK.queryTask(dependence);\\
	\If{$flag$ $\neq$ 'processing'}{
		\KwRet{err};
	}
	$t_j$ $\leftarrow$ \{$tN$, manager, $bT$, $eT$, $flag$, info, dependence, $cT$, $pN$\};\\
	
	\If{$t_j$ $\notin$ $T^{n}_{i}$ || $t_j$ does not depend on $T'$}{
		\KwRet{err};
	}
	res $\leftarrow$ Call SDK.assignTask($t_j$);\\
	\If{res is successful}{
		Send the sign of success to the client, and the client requests the 'process\_gantt' module again.
	}
\end{algorithm}

\textbf{Assign:} The task scheduling process is assigned by the person in charge of the project $p$. As shown in Algorithm \ref{al:a4}, manager $u_i$ sends the task data to the server, and the server runs the "assign\_task" module. The assign task scheduling sets the value of $completedTime$ as $null$ and attribute $projectName$ as the name of project $p$ to construct task data structure $t_j$. Then, it retrieves $t_j$'s dependent task set, $T^{'}$, through SDK and attribute $dependence$. When checking the legitimacy of $t_j$, the assign task scheduling first checks whether the $flag$ is $processing$, then checks whether the start and end times of $t_j$ are within the scope of $p$, and finally checks whether $t_j$ conforms to the dependency set $T^{'}$. If $t_j$ is legal, the $assignTask()$ method of the chaincode is called to write $t_j$ to the state database and return the execution result to the client. Finally, if the execution is successful, the client requests the server to execute the "process\_gantt" module again.

\begin{algorithm}[ht]
	\caption{Adjust completion for $t_j$}
	\label{al:a5}
	\KwIn{$tN$, $cT$, $u_i$}
	\KwOut{}
	
	The server runs the 'setCompletedTime' module;\\
	$t_j$ $\leftarrow$ Call SDK.queryTask($tN$);\\
	$p$ \ $\leftarrow$ Call SDK.queryProject($t_j$.projectName);\\
	\If{$t_j$.manager $\neq$ $u_i$ || $p$.manager $\neq$ $u_i$}{
		\KwRet{err};
	}
	\If{cT $\notin$ ($t_j$.beginTime, $t_j$.endTime]}{
		\KwRet{err};
	}
	$t_j$.completedTime $\leftarrow$ $cT$;\\
	\If{cT == $t_j$.endTime}{
		$t_j$.flag = '$done$';\\
		\If{$t.flag$ == $done$, $\forall t\in p$}{
			$p.flag$ = '$done$';
		}
	}
	res $\leftarrow$ Call SDK.changeTask($tN$, 'changeInfo', $t_j$);\\
	\If{res is successful}{
		Send the sign of success to the client, and the client requests the 'process\_gantt' module again.
	}
\end{algorithm}

\textbf{Change:} The system proposed Fabric-GC not only provides project participants with shared task scheduling data for the entire project but also provides shared completion of the project during execution. Algorithm \ref{al:a5} outlines the process of adjusting the completion of a specified task scheduling. The client member $u_i$ specifies the $\{taskName,completedTime\}$ to send to the server, which runs the 'setCompletedTime' module. Next, the SDK obtains the task scheduling $t_j$ and the project $p$ to which $t_j$ belongs. If $u_i$ is not the person in charge of $t_j$ and does not belong to the manager of $p$, this means that $u_i$ does not have the permission to modify $t_j$ and the system returns a permission error. Before any modification, the system ensures that the completion degree $completedTime$ falls within the starting and ending range of $t_j$. After setting the completion degree of $t_j$, if $completedTime==t_j.endTime$, the system sets $t_j.done=done$ and checks whether $\forall t.flag=done,\ t\in p$ holds and the entire project has been completed. After the successful execution, the server will return the execution result to the client, and the client will request the server to execute the "process\_gantt" module again.

\textbf{View:} This module visualizes all task scheduling sets as a Gantt chart. Besides, after getting all the task name sets $tasks$ of $p$, this module calls the $getTasks()$ method provided by the server. Through iterating $tasks$, the chaincode $queryTask()$ method is called to retrieve the entire task collection. Finally, this module calls the $ganttView()$ to visualize the task set as a Gantt chart.

\section{Experiment and Comparison}
\label{sec:experiment}
This section describes the experimental process and the results achieved by evaluating the functions and performance of the proposed Fabric-GC system. The creation process of the system and the implementation of cross-organizational project management based on the system are depicted and followed by comparative experiments and performance results and analysis.

\subsection{Design, Implementation, and Evaluation of fabric-GC}
The experiments are divided into two parts. The former describes the network structure and project structure of Fabric-GC. At the same time, the latter introduces the operation steps of Fabric-GC, including how to create projects in a multi-organization environment, task scheduling, and Gantt chart visualization.

There are six docker containers in Fabric-GC runtime that constitute the Blockchain network, as shown in Table \ref{tab:nodes}.

\begin{table}[ht]
	\centering
	\caption{Fabirc-GC nodes.}
	\label{tab:nodes}
	\begin{tabular}{c|c|c}
		\toprule
		\textbf{Node name} & \textbf{Description} & \textbf{Number} \\
		\midrule
		fabric-couchdb & database node & 4 \\
		fabric-ca & CA node & 2 \\
		fabric-peer & peer node & 4 \\
		fabric-orderer & orderer node & 1 \\
		fabric-tools & cli node & 1 \\
		fabric-gantt/chaincode & chaincode node & 2 \\
		\bottomrule
	\end{tabular}
\end{table}

The complete project structure consists of four parts.

\begin{enumerate}
	\item \textbf{bin:} Binary tool directory. It is mainly used to generate certificates, block configuration, channel configuration, and other files.
	\item \textbf{chaincode:} The directory where the chaincode is stored.
	\item \textbf{client:} The main directory of the project. Save the network startup script, chaincode installation script, server-side source code.
	\item \textbf{network:} Network configuration file directory. It includes a docker container configuration file, block configuration file, and certificate generation file.
\end{enumerate}

The project initialization steps are as follows.

\textbf{Step 1:} Run the script code, \textit{start.sh}. This script first removes the volume nodes and data that have been started. Next, it generates the certificate file, block configuration, channel configuration, and other files required by the startup container. The container is started, the channel initialized, and each peer node is added to the channel. Lastly, the chaincode is installed and initialized.

\textbf{Step 2:} Create administrator accounts for $Org1$ and $Org2$, and account information not stored in the Blockchain.

\textbf{Step 3:} Start the server process to receive client requests.

\subsubsection{Evaluation of Fabric-GC in the multi-organizational environment}
This experiment assumes that the project manager is $user1$ and belongs to organization $Org1$. The project $project1$ is divided into six tasks assigned to different participants, where $user2$ and $user3$ belong to organization $Org1$, while $user4$, $user5$, $user6$ and $user7$ belong to organization $Org2$.

\begin{table}[ht]
	\centering
	\caption{The task set of $project1$.}
	\label{tab:taskSet}
	\begin{tabular}{|c|c|c|c|c|}
		\hline
		\textbf{Task Name} & \textbf{Principal} & \textbf{Organization} & \textbf{beginTime} & \textbf{endTime}  \\
		\hline
		task1 & user2 & Org1 & 2020.11.15 & 2020.11.28 \\
		task2 & user3 & Org1 & 2020.11.29 & 2020.12.05 \\
		task3 & user4 & Org2 & 2020.12.06 & 2020.12.10 \\
		task4 & user5 & Org2 & 2020.12.11 & 2020.12.15 \\
		task5 & user6 & Org2 & 2020.11.29 & 2020.12.10 \\
		task6 & user7 & Org2 & 2020.12.16 & 2020.12.31\\
		\hline
	\end{tabular}
\end{table}

Each project member is registered in the system through its terminal. $user1$ logs into the system and creates project $project1$ at first hand, and then assign task scheduling for members according to Table \ref{tab:taskSet}. After that, each member logs into Fabric-GC. After successful login, the project list of all projects that the member participates in is displayed. Project $project1$ is queried by $user4$, indicating that data sharing is successful.

\begin{figure}[ht]
	\centering
	\includegraphics[width=\linewidth]{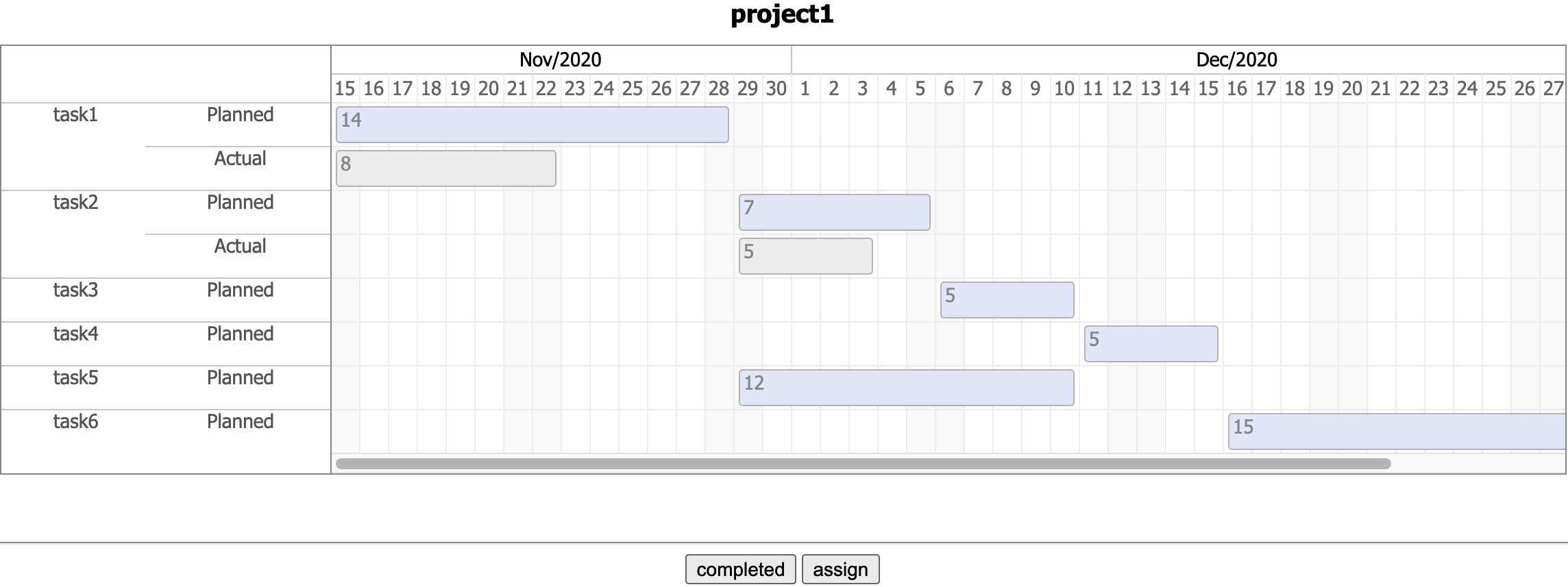}
	\caption{The interface of Gantt chart}
	\label{fig:gantt}
\end{figure}

\begin{figure}[ht]
	\centering
	\includegraphics[width=\linewidth]{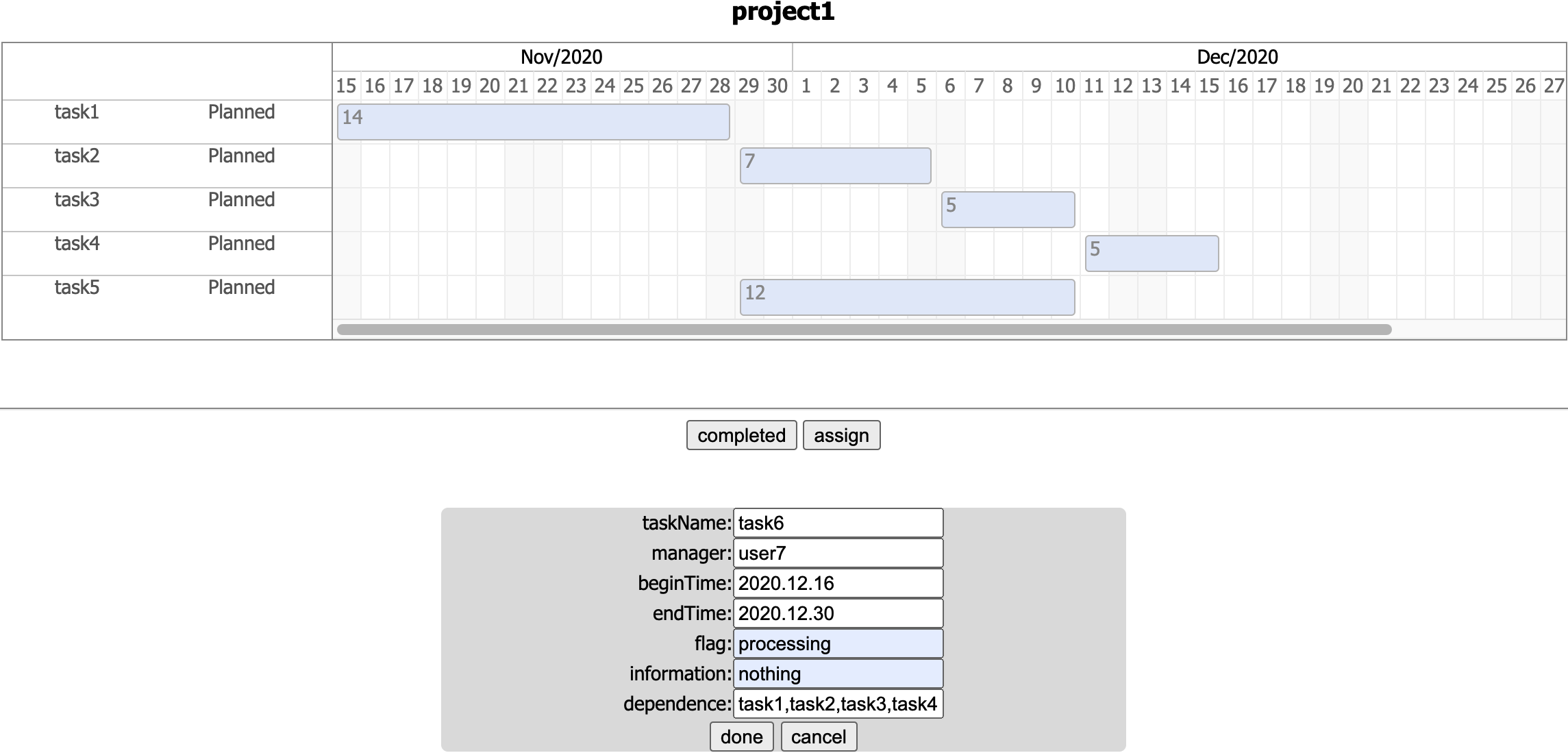}
	\caption{Assign a task schedule.}
	\label{fig:assign}
\end{figure}

The Gantt chart represented by $project1$ is shown in Fig~\ref{fig:gantt}, where the blue bar chart represents the planned duration, while the gray bar chart represents the completed duration. The status of a project implementation can be seen from the graph. That is, the Blockchain system ensures that all members participating in the project can obtain the latest status information.

\begin{figure}[ht]
	\centering
	\includegraphics[width=\linewidth]{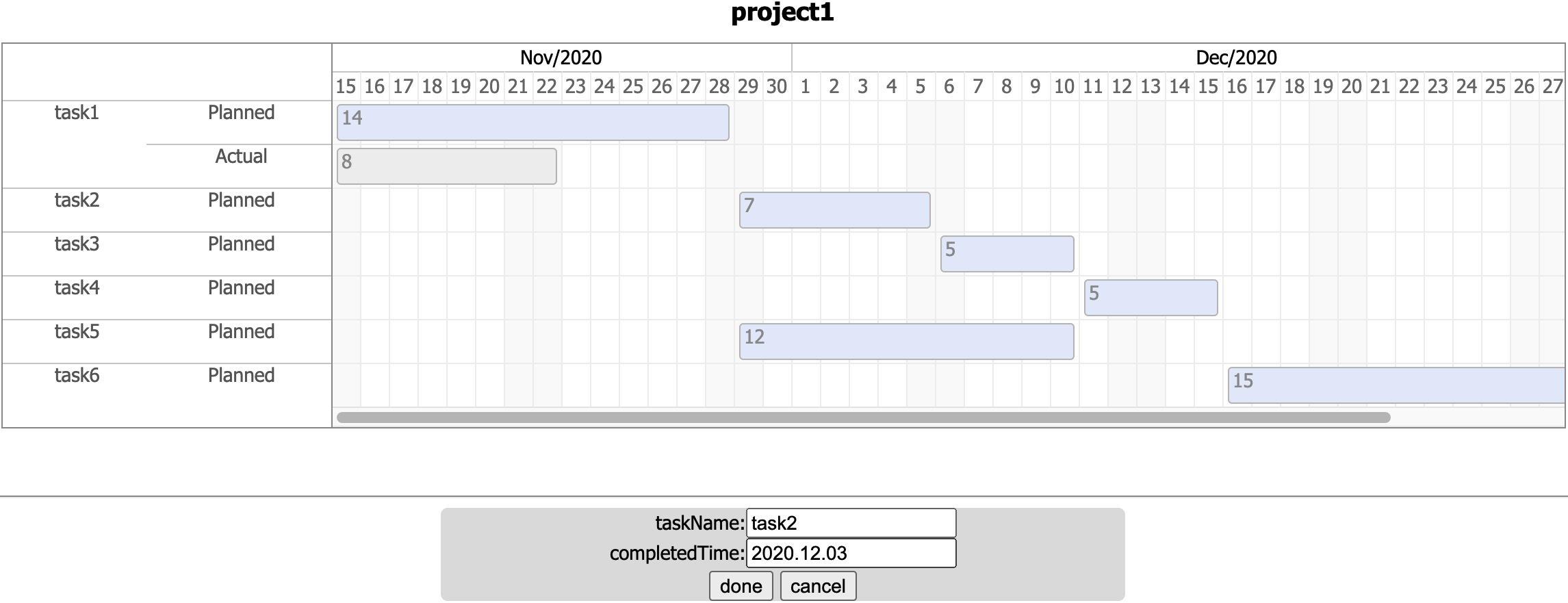}
	\caption{The process of completion setting.}
	\label{fig:setCom}
\end{figure}

To assign task scheduling to $project1$, as shown in Fig~\ref{fig:assign}, $user1$ can select the 'assign' button to pop up the dialog box and fill in the information of the task to be assigned. When $user3$ needs to feedback the completion progress of $task2$, click the 'completed' button, as shown in Fig~\ref{fig:setCom}, and specify the task name and completed time. After $user1$ receives the feedback information, he can adjust the project in real-time according to the completion status and repeat the above process so that the project Manager $user1$ can always grasp the project's overall implementation to achieve the goal of the project saving resource cost and improving execution efficiency.

\subsection{Result and Comparison}

\begin{figure}[ht]
	\centering
	\includegraphics[width=0.7\linewidth]{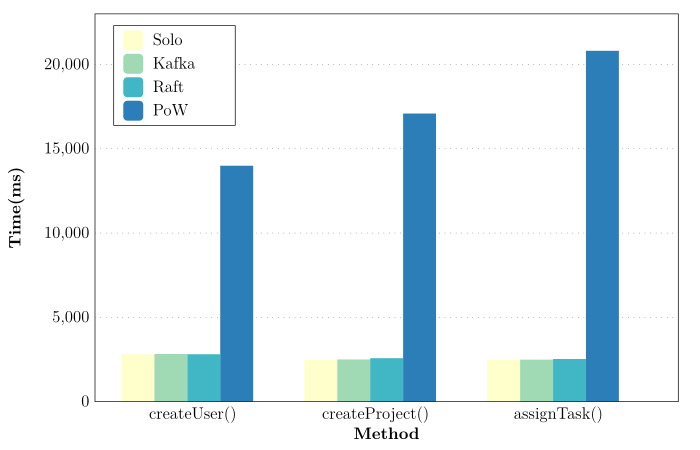}
	\caption{Execution time of write operations under Solo, Kafka, Raft and PoW.}
	\label{fig:Consensus}
\end{figure}

To test the performance of the Fabric-GC, we compare the execution time when the project data is stored and choose \emph{tape} \cite{tape} for the TPS (Transaction Per Second) throughput testing of the chaincode. We remark that the performance bottleneck of Blockchain is mainly in the consensus mechanism, and different consensus mechanisms impact the data synchronization rate between nodes. Since the nodes of the public chain do not trust each other, the \emph{PoW (Proof-of-Work)} consensus algorithm is required to achieve data synchronization. In this case, the nodes' arithmetic power is used to mine for packing rights, and its execution is inefficient. As shown in Fig~\ref{fig:Consensus}, the execution time of PoW impacts more than $10$ seconds on the write operations $createUser()$, $createProject()$, and $assignTask()$. On the other hand, if the nodes of the consortium chain trust each other, the data synchronization time is about $2.5$ seconds under the consensus algorithm (\emph{Solo}, \emph{Kafka}, and \emph{Raft}).

\begin{figure}[ht]
	\centering
	\includegraphics[width=\textwidth]{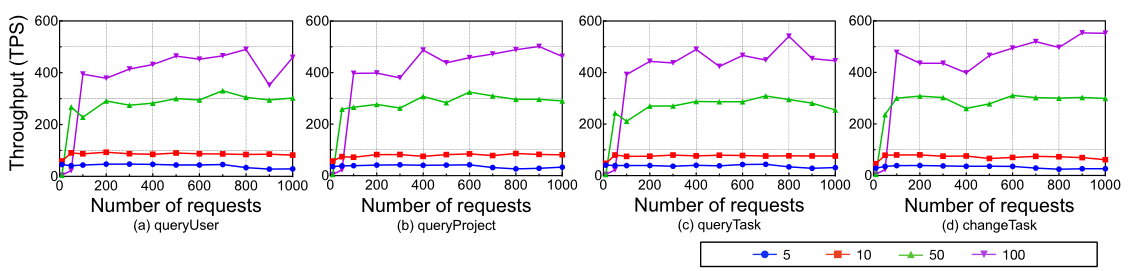}
	\caption{Throughput under different $Transactions$. $Transactions$ represents the number of transactions in a block.}
	\label{fig:tps}
\end{figure}

Several methods with high request in Fabric-GC use read operations, e.g., $queryUser()$, $queryProject()$, $queryTask()$ and $changeTask()$. Fig~\ref{fig:tps}(a)-\ref{fig:tps}(d) show TPS under different $Transactions$. Under the same number of transactions, the TPS of the four methods has little difference. This aspect shows that Fabric processes various transactions in the same process. In addition, we can observe that, when $Transactions=100$, and the number of requests is more than $100$, the throughput of the system can reach ranging $400$ to $500$. Besides, when $Transactions=5$, the throughput of the system is around $50$. The smaller the number of transactions, the more blocks generated in the same time period. The system consumes too many resources for the packaging and verification of blocks.

\begin{figure}[ht]
	\centering
	\includegraphics[width=0.7\linewidth]{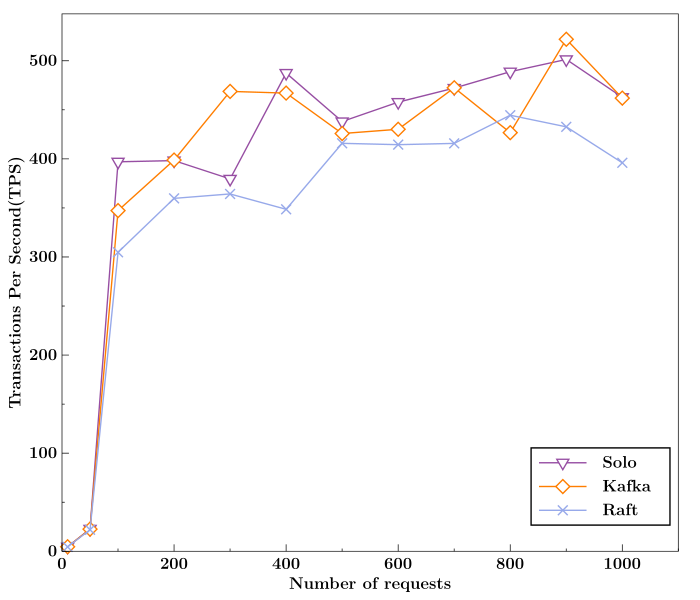}
	\caption{Throughput of $queryProject()$ under different consensus algorithms when $Transactions=100$.}
	\label{fig:com}
\end{figure}

Fig~\ref{fig:com} shows the throughput curve of $queryProject()$ method under different consensus algorithms when the block size is set to $100$MB. More precisely, with the Solo and Kafka consensus mechanism, the difference between their TPS is not significant. When the number of requests exceeds $400$, the system throughput can be kept from $400$ and $500$. The throughput of Raft is lower than the other two. Therefore, Kafka's consensus should be selected in the production environment as far as possible to ensure high performance and high fault tolerance.

The comparative experiments show that fabric-GC can maintain high throughput under large-scale requests and adapt to Blockchain networks under different consensus algorithms with relatively stable performance output.

\section{Conclusions and Future Work} 
\label{sec:conclusion}
Multi-party project cooperation is a standard practice, widely used in scientific R \& D, industrial production, software development, supply chain \cite{9047881} among several other fields. Indeed, the collaboration between organizations and individuals with different technologies improves the rate of the success of complex projects \cite{7851142}. Nevertheless, cross-organizational projects pose difficulties for project managers in managing task scheduling and progress feedback that relies on timely information sharing \cite{YANG2020103276}. The independence and heterogeneity among participating organizations can make data sharing difficult. Moreover, since traditional data sharing relies on third-party organizations (e.g., cloud, specialized service provider, transcription services, call center services, consulting), the privacy and security of data cannot be guaranteed \cite{9142202}.

In this article, we propose a Blockchain-based Gantt chart system, named \emph{Fabric-GC}. The proposed system mitigates the difficulties arising from human resource management and information transfer in multi-organizational project cooperation scenarios. In this proposed research, the Blockchain eliminates the heterogeneity between different partners, enabling them to maintain and manage the same project jointly. In detail, with the support of smart contracts, the project manager can communicate the Gantt chart schedule to the participants across the organization through Fabric-GC. Therefore, the participants can achieve real-time feedback on the project progress, and the project manager can make timely adjustments to the project schedule. Experimental results show that the proposed system can deal with large-scale data request scenarios while maintaining stable performance under different consensus mechanisms.

As future research directions, we intend to work on the following list of items:

\begin{enumerate}
	\item The performance testing and evaluation of \emph{Fabric-GC} have been conducted in a single machine environment. As future work, we plan to consider a distributed environment containing multiple nodes for testing and verifying the performance of our proposal;
	\item In project management, not only time cost needs to be considered, but also resource allocation, among other issues and costs. Additional features in this regard are under investigation and will be included in future extensions of \emph{Fabric-GC};
	\item \emph{Fabric-GC} cannot store a large amount of data. Interconnecting \emph{Fabric-GC} with distributed storage systems such as the \emph{Interplanetary File System (IPFS)} will be considered to overcome this limitation. In this way, \emph{Fabric-GC} will also enable the sharing of project-related resources, including files, video, audio, and other resources.
	\item The management of participants will be enhanced to realize the evaluation of participants' capabilities. In this way, project managers can make more effective and reasonable project planning and execution.
\end{enumerate}


\section*{Acknowledgement}
This research is supported by the National Natural Science Foundation of China under Grant 61873160, Grant 61672338, and the Natural Science Foundation of Shanghai under Grant 21ZR1426500.

\section*{Declaration of interests}
The authors declare that they have no known competing financial interests or personal relationships
that could have appeared to influence the work reported in this paper.

\bibliographystyle{splncs03}
\bibliography{reference}

\par\noindent 
\parbox[t]{\linewidth}{
\noindent\parpic{\includegraphics[height=1.5in,width=1in,clip,keepaspectratio]{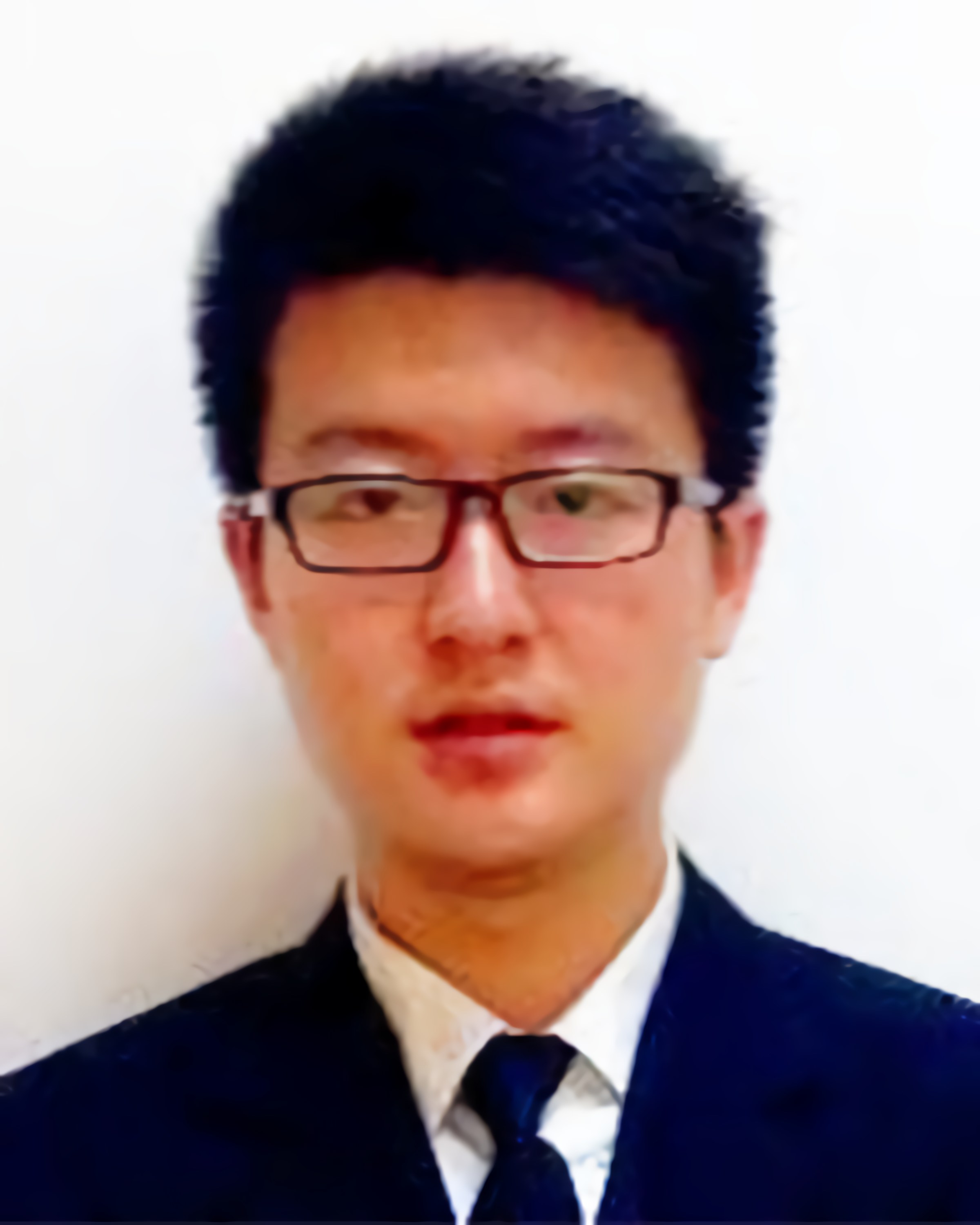}}
\noindent {\bf Dun Li}\
received the B.S. degree in Human Resource Management from the Huaqiao University, Quanzhou, China, in 2013,  and the M.S. degree in Finance from the Macau University of Science and Technology, Macau, China, in 2015. 
He is currently doing his Ph.D. degree in Information Management and Information Systems at Shanghai Maritime University.
His research interests mainly include smart finance, big data, machine learning, IoT, and blockchain.}
\vspace{4\baselineskip}

\par\noindent 
\parbox[t]{\linewidth}{
\noindent\parpic{\includegraphics[height=1.5in,width=1in,clip,keepaspectratio]{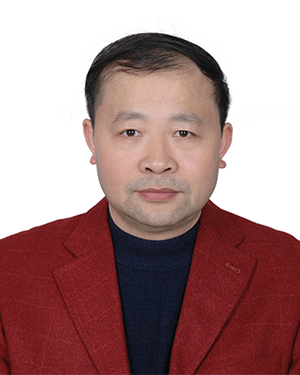}}
\noindent {\bf Dezhi Han}\
received the BS degree from Hefei University of Technology, Hefei, China, the MS  degree and PhD degree from Huazhong University of Science and Technology, Wuhan, China. 
He is currently a professor of computer science and engineering at Shanghai Maritime University. His specific interests include storage architecture, blockchain technology, cloud computing security and cloud storage security technology. }
\vspace{4\baselineskip}

\par\noindent 
\parbox[t]{\linewidth}{
\noindent\parpic{\includegraphics[height=1.5in,width=1in,clip,keepaspectratio]{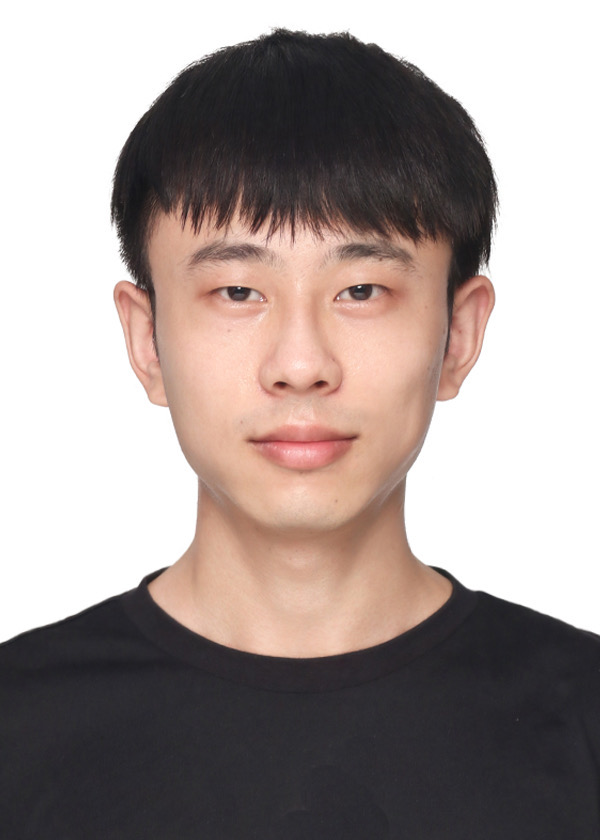}}
\noindent {\bf Benhui Xia}\
received the B.S. degree from China University of Mining and Technology, where he is currently pursuing the M.S. degree with Shanghai Maritime University. His main research interests include network security, cloud computing, distributed computing and blockchain.}
\vspace{4\baselineskip}

\par\noindent 
\parbox[t]{\linewidth}{
\noindent\parpic{\includegraphics[height=1.5in,width=1in,clip,keepaspectratio]{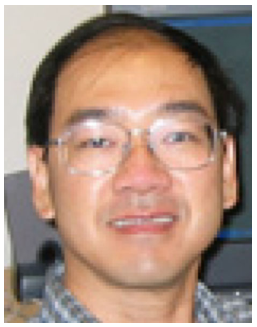}}
\noindent {\bf Tien-Hsiung Weng}\
is currently a professor at the Department of Computer Science and Information Engineering at Providence University, Taichung, Taiwan. 
He received a Ph.D. in Computer Science from the University of Houston, USA. 
His research interests include parallel programming models, performance measurement, and compiler analysis for code improvement.}
\vspace{4\baselineskip}

\par\noindent 
\parbox[t]{\linewidth}{
\noindent\parpic{\includegraphics[height=1.5in,width=1in,clip,keepaspectratio]{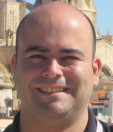}}
\noindent {\bf Arcangelo Castiglione}\
received a Ph.D. degree in Computer Science from the University of Salerno, Italy. He is a tenure-track assistant professor at the Department of Computer Science, University of Salerno (Italy). His research mainly focuses on cryptography, network security, data protection, digital watermarking, and automotive security. He is an Associate Editor for several Scopus-Indexed journals, and he has been Guest Editor for several Special Issues and Volume Editor for Lecture Notes in Computer Science (Springer). He has been involved in several organizational roles (steering committee member, program chair, publicity chair, etc.) for many international conferences. He has been a reviewer for several top-ranked scientific journals and conferences. He has been appointed as a member of the IEEE Technical Committee on Secure and Dependable Measurement. He is a founding member of the IEEE TEMS Technical Committee (TC) on blockchain and Distributed Ledger Technologies. }
\vspace{4\baselineskip}

\par\noindent 
\parbox[t]{\linewidth}{
\noindent\parpic{\includegraphics[height=1.5in,width=1in,clip,keepaspectratio]{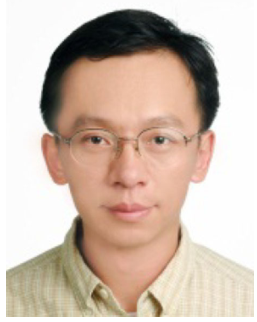}}
\noindent {\bf Kuan-Ching Li}\
is currently appointed as a professor in the Dept. of Computer Science and Information Engineering (CSIE) at Providence University, Taiwan, where he also serves as the Director of the High-Performance Computing and Networking Center. Besides the publication of articles in renowned journals and conferences, he is co-author or co-editor of more than 40 books published by Taylor \& Francis, Springer, IGI Global and McGraw-Hill. He is a Fellow of IET and a senior member of the IEEE. Professor Li's research interests include parallel and distributed computing, Big Data, and emerging technologies.}
\vspace{4\baselineskip}
\end{document}